\documentclass[conference]{IEEEtran}
\IEEEoverridecommandlockouts
\usepackage{cite}
\usepackage{amsmath,amssymb,amsfonts}
\usepackage{algorithmic}
\usepackage{graphicx}
\usepackage{textcomp}
\usepackage{xcolor}
\usepackage{bm}
\usepackage{mathptmx,mathtools}
\usepackage{soul}
\DeclareSymbolFont{Xlargesymbols}{OMX}{cmex}{m}{n}
\DeclareMathSymbol{\Xsum}{\mathop}{Xlargesymbols}{80}

\newcommand{\lyxdot}{.}

\def\BibTeX{{\rm B\kern-.05em{\sc i\kern-.025em b}\kern-.08em
    T\kern-.1667em\lower.7ex\hbox{E}\kern-.125emX}}
\begin{document}

\title{An unsupervised machine-learning checkpoint-restart algorithm using Gaussian mixtures for particle-in-cell simulations*\\
{\footnotesize \textsuperscript{}}
\thanks{This work was supported by the U.S. Department of Energy, Office of Science, Office of Applied Scientific Computing Research (ASCR), both by the EXPRESS (2016-17) and SciDAC (2018-21) programs.}
}
\author{\IEEEauthorblockN{Guangye Chen}
\IEEEauthorblockA{\textit{Theoretical Division} \\
\textit{Los Alamos National Laboratory}\\
Los Alamos, NM, USA \\
gchen@lanl.gov}
\and
\IEEEauthorblockN{Luis Chac\'on}
\IEEEauthorblockA{\textit{Theoretical Division} \\
\textit{Los Alamos National Laboratory}\\
Los Alamos, NM, USA \\
chacon@lanl.gov}
\and
\IEEEauthorblockN{Truong B. Nguyen}
\IEEEauthorblockA{\textit{Theoretical Division} \\
\textit{Los Alamos National Laboratory}\\
Los Alamos, NM, USA \\
tbnguyen@lanl.gov}
}

\maketitle

\begin{abstract}
We propose an unsupervised machine-learning checkpoint-restart (CR) lossy algorithm for particle-in-cell (PIC) algorithms using Gaussian mixtures (GM). The algorithm features a particle compression stage and a particle reconstruction stage, where a continuum particle distribution function is constructed and resampled, respectively. To guarantee fidelity of the CR process, we ensure the exact preservation of charge, momentum, and energy for both compression and reconstruction stages, everywhere on the mesh. We also ensure the preservation of Gauss' law after particle reconstruction. As a result, the GM CR algorithm is shown to provide a clean, conservative restart capability while potentially affording orders of magnitude savings in input/output requirements. We demonstrate the algorithm using a recently developed exactly energy- and charge-conserving PIC algorithm on physical problems of interest, with compression factors $\gtrsim75$ with no appreciable impact on the quality of the restarted dynamics.
\end{abstract}


\section{Introduction}
Resiliency, data locality, and asynchrony are key major challenges
facing the practical use of exascale computing for scientific applications.
Because of extreme concurrency, very large system scale, and complex
memory hierarchies, hardware failures (both ``soft'' and ``hard'')
are expected to become more frequent towards and beyond exascale.
Currently, supercomputers with 100 billion transistors/node, thousands
of nodes, and 10M cores are operational (e.g., Summit and Sierra \cite{kahle2019}).
The very large total number of components will lead to frequent failures,
even though the mean time between failures (MTBF) for the individual
components may be large. For instance, while the MTBF of a CPU can
be months to years \cite{nightingale2011cycles}, that of current
supercomputers can be a few hours or less \cite{liu2018large,rojas2019analyzing}.
With billion-core parallelism at exascale, the MTBF has been projected
to be within (or even significantly below) one hour \cite{dauwe2017analysis,miao2018energy}.
Therefore, it is important to enable efficient strategies that allow
software and algorithms to perform in a frequently interrupted environment.

Particle-based
simulations at the extreme scale are particularly challenged by the
input/output (IO) requirements of storing billions to trillions of
particles, as is already the case for conventional checkpoint/restart (CR) in the leading-class supercomputers. Here,
we focus on reducing the storage requirement by an unsupervised machine-learning technique for kinetic plasma simulations using the particle-in-cell (PIC) method \cite{birdsall2004plasma,hockney1988computer}. Compression of particle data
is performed per spatial cell by construction of a velocity distribution
function (VDF) with a Gaussian mixture \cite{mclachlan2004finite},
based on a penalized maximum-likelihood-estimation (PMLE) approach
\cite{figueiredo2000unsupervised}. The resulting optimization problem
is solved by an adaptive Expectation-Maximization (EM) algorithm \cite{dempster1977maximum,figueiredo2000unsupervised},
which can automatically search for the optimal number of Gaussian
components satisfying a generalized ``minimum-message-length (MML)''
Bayesian Information Criterion \cite{wallace2005statistical}. The
method can be formulated to conserve up to second moments exactly
\cite{behboodian1970mixture}, and can be solved efficiently
\cite{nguyen2020adaptive}. Particle-data is reconstructed (also
locally per cell) by  Monte Carlo sampling of the VDF
with a simple moment-matching projection technique to conserve momentum
and energy exactly \cite{lemons2009small}. \textcolor{black}{Particle
spatial positions within a given cell are re-initialized randomly
assuming that the plasma is uniform within a cell (other models are possible).} Both
compression and reconstruction operations are local in configuration
space (i.e., each computational cell features an \emph{independent}
VDF reconstruction process) and done \emph{in-situ} for cells with
enough particles (e.g., more than 10), and only Gaussian parameters
are checkpointed.

\section{Methodology}
A GM is defined as a convex combination of $K$ Gaussian distributions:
\begin{equation}
p(\mathbf{x})=\Xsum_{k=1}^{K}\omega_{k}f_{k}(\mathbf{x}),\label{eq:mix-dist}
\end{equation}
where each Gaussian $f_{k}$ is weighted by $\omega_{k}$ with $\Xsum_{k}\omega_{k}=1$
and $\omega_{k}>0$. The Gaussian distribution is defined as 
\begin{equation}
f_{k}(\mathbf{x})=\frac{1}{\sqrt{(2\pi)^{D}|{\boldsymbol{\Sigma}}^{}_{k}|}}e^{-(\mathbf{x}-\boldsymbol{\mu}_{k})^{T}{\boldsymbol{\Sigma}}^{-1}_{k}(\mathbf{x}-\boldsymbol{\mu}_{k})/2},
\end{equation}
where $\boldsymbol{\mu}$ is a $D$-dimensional mean vector, ${\boldsymbol{\Sigma}}$
is a $D\times D$ covariance matrix, and $|{\boldsymbol{\Sigma}}|$
is the determinant of ${\boldsymbol{\Sigma}}$.

The goal is to estimate both the parameters $\bm{\theta}_{k}\equiv\{\boldsymbol{\omega},\boldsymbol{\mu},{\boldsymbol{\Sigma}}\}_{k}$
of each Gaussian as well as the number of Gaussians, $K$,
given $N$ independent samples $\mathbf{X}$$=(\mathbf{x}_{1}...\mathbf{x}_{N}$). This is accomplished in the framework of the Bayesian information
criterion by maximizing a penalized likelihood function \cite{figueiredo2000unsupervised}, 
\begin{equation}
L(\bm{\theta})=\Xsum_{p=1}^{N}\alpha_{p}\mathrm{ln}\left[\Xsum_{k=1}^{K}\omega_{k}f_{k}(\mathbf{v}_{p}|\boldsymbol{\mu}_{k},{\boldsymbol{\Sigma}}_{k})\right]-\frac{d}{2}\mathrm{ln}N-\frac{T}{2}\Xsum_{k=1}^{K}\mathrm{ln}(\omega_{k}),\label{eq:lnL-GM}
\end{equation}
{where $\alpha_{p}$ is the particle weight, $d$ is the total number of Gaussian parameters,
$N$ is the total number of samples, and $T=D(D+3)/2$ is the number
of parameters specifying each Gaussian component.} By setting the partial derivatives with respect to Gaussian parameters to zeros, we obtain a set of likelihood equations which can be solved iteratively by an adaptive expectation-maximization (EM) algorithm \cite{figueiredo2000unsupervised}.

It is worth mentioning that the EM-GM algorithm based on standard (\emph{unpenalized})
maximum-likelihood-estimator conserves up to second moments of the sample particles
\emph{exactly} at every iteration, i.e., the mass, mean, and variance
of the mixture coincide with those of the sample particles. The penalization terms in adaptive EM break this property. To recover strict conservation, we perform an additional standard EM iteration after the adaptive EM is converged \cite{chen2020unsupervised}.

We only store Gaussian parameters to disk to checkpoint the particles. After restart, we begin by sampling the Gaussian mixture in velocity space using standard Monte Carlo sampling. Next, we perform a Lemon's moment correction \cite{lemons2009small,chen2020unsupervised} to the sampled particles to recover momentum and energy conservation. Subsequently, we enforce Gauss's law by performing a global mass-matrix solve \cite{burgess1992mass,chen2020unsupervised}, ensuring that the charge densities on grid-points are identical to those of the pre-checkpointed system.

\section{Results}
\subsection{Correctness test}

\begin{figure}
\begin{centering}
\includegraphics[width=1\columnwidth,trim={ 0 0 0 0 },clip]{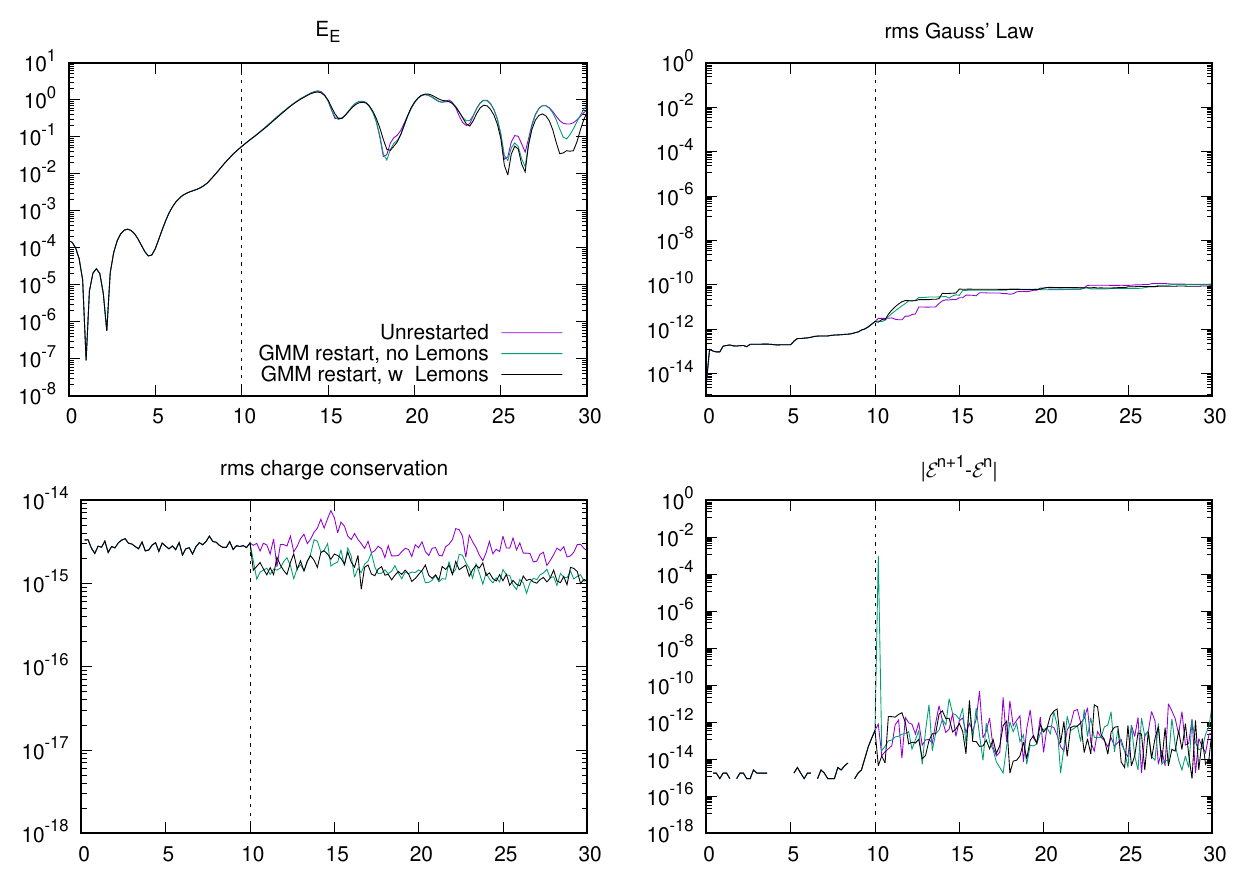}
\par\end{centering}
\caption{\label{fig:Two-stream-instability-restarted}Two-stream instability:
Semi-log-scale time history of the electric field energy $E_{E}$
(top-left), the rms of Gauss' law residual over the whole mesh (top-right),
the rms of the residual of the charge conservation equation (bottom-left),
and the change of total energy between subsequent time steps (bottom-right).
The simulations are obtained without restart, and with GM restart
at $t=10$ (in normalized units) with and without Lemons moment matching.}
\end{figure}

We demonstrate the proposed CR algorithm using a prototypical test problem, the 1D-1V
electrostatic two-stream instability (electromagnetic tests are provided elsewhere\cite{chen2020unsupervised}). Simulations were performed with the DPIC code,
based on an implicit, charge and energy conserving multidimensional electromagnetic Darwin PIC algorithm \cite{chen2015multi}.
The two-stream instability is an electrostatic
instability in which two counter-streaming particle beams exchange
kinetic and electrostatic energy, and as a result tangle up in to
a vortex in phase space \cite{roberts1967nonlinear}. The simulation
is performed for $L=2\pi$ (domain size, in Debye length units), $v_{b}=\sqrt{3}/2$
(beam speed, in electron thermal speed units), $N_{x}=32$ (number
of cells), $N_{p}=156$ (number of particles per cell), $\Delta t=0.2$
(time step in inverse plasma frequency units), with periodic boundary
conditions. Figure \ref{fig:Two-stream-instability-restarted} shows
 the electric-field
energy, errors in Gauss' law and charge conservation, and the change in total energy between subsequent timesteps (which should be zero for strict conservation). The plot compares the
unrestarted run with two GM-restarted ones at $t=10$ (mid/late linear
stage), with and without Lemons' moment correction. The results show
exact conservation of charge for all cases, also for energy except
for the case without Lemons correction (which has
a large energy conservation error right after restart), and excellent
preservation of Gauss' law (commensurately with the nonlinear tolerance).
They also show excellent agreement in the temporal evolution of the
electrostatic field energy for all cases. For this run, the GM algorithm
is started with 8 Gaussian's per cell, resulting in an average number
of Gaussians per cell of 2, and therefore to an average compression
ratio  \textcolor{black}{(defined as the ratio of original and compressed data sizes)} of about 75. \textcolor{black}{Similar compression ratios have been demonstrated for multidimensional electromagnetic applications as well \cite{chen2020unsupervised}.}

\begin{figure}
\begin{centering}
\includegraphics[width=0.35\columnwidth,angle=-90,trim={ 0.2in 0 0 0 },clip]{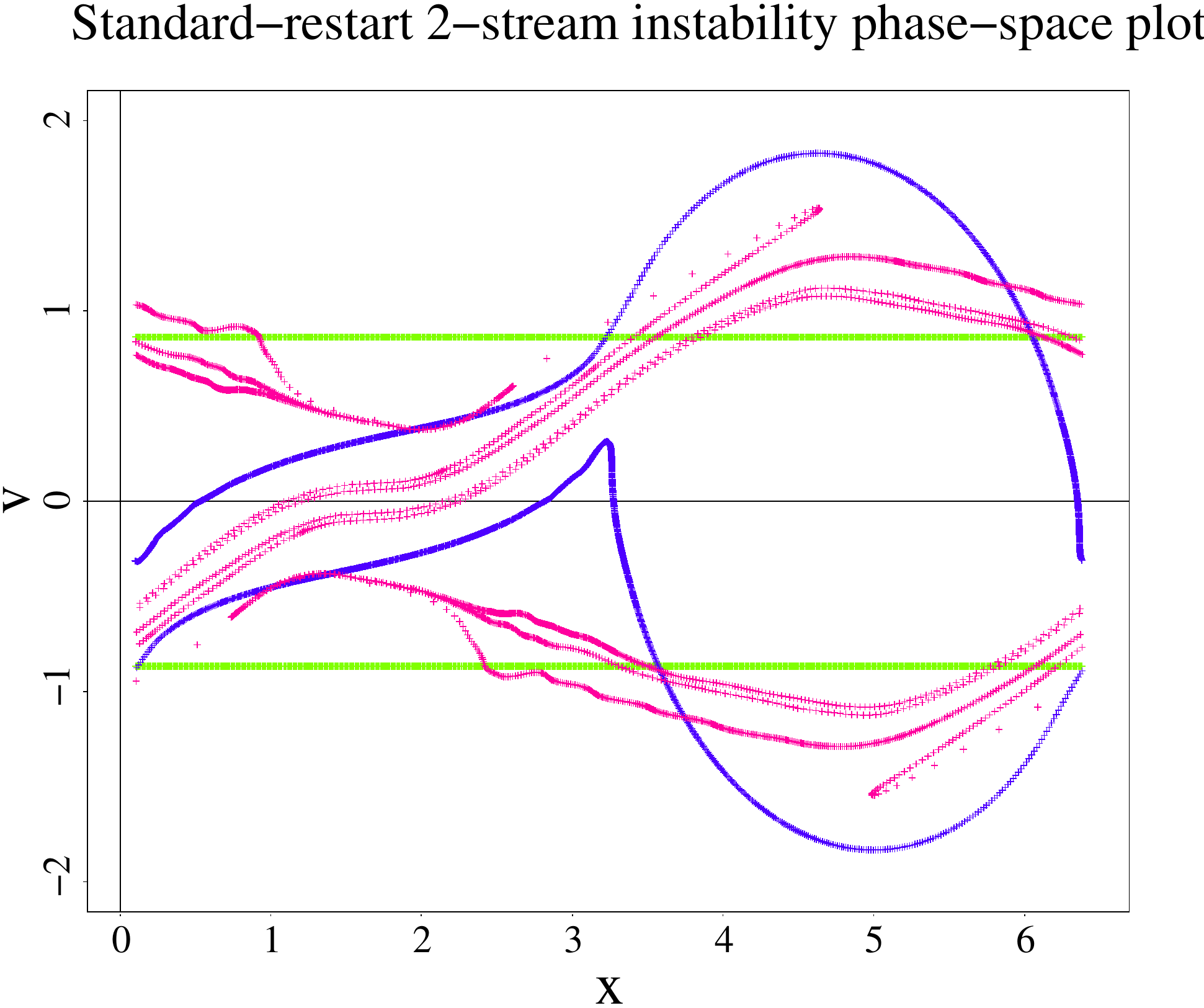}\qquad{}\includegraphics[width=0.35\columnwidth,angle=-90,trim={ 0.2in 0 0 0 },clip]{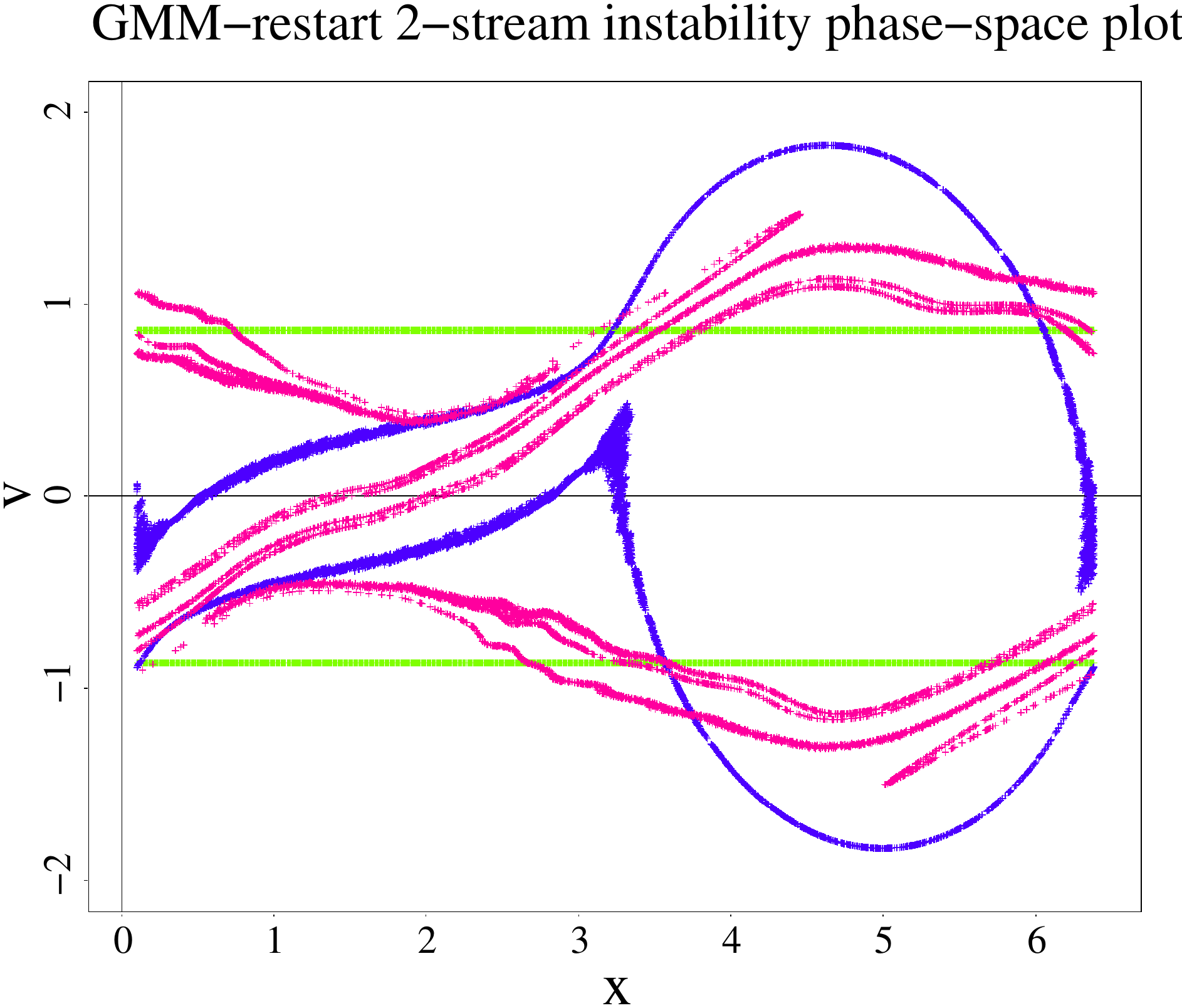}
\par\end{centering}
\caption{\label{fig:2-stream-phase-space}Two-stream instability: Phase-space
(x-v) comparison at three different times (green: $t=0$, blue: $t=14.0$,
red: $t=19.4$) between the unrestarted case (left) and the GM-restarted
one (right).}
\end{figure}

A comparison of 1D-1V phase-space plots between unrestarted (left)
and GM-restarted (right) runs is shown in Fig. \ref{fig:2-stream-phase-space}.
It demonstrates that the GM algorithm (right) is capable of
capturing fine phase-space features present in the unrestarted case
(left). {\color{black}The beam-spread present in the particles originates in the random spatial re-initialization per cell in the GM-restart, and is noticeable in this case because the phase space initial condition for this problem is unusually sharp (it is a delta function in velocity). Subsequent time evolution thermalizes the beam, and makes the spread inconsequential from a dynamical perspective. Most plasma applications feature some thermal spread, which makes this issue moot.

In lossless compression, all the particle velocities and positions (or, equivalently, all independent moments of the original particle distribution) would have to be preserved. Our approach conserves exactly up to second moments. The underlying penalized maximum likelihood principle ensures that the main structures of the distribution are captured by the Gaussians. Higher-order moments are captured by convergence, with the error scaling as the inverse square-root of the number of particles if random sampling is employed, or better if quasirandom sampling is employed~\cite{lemieux2009monte}. The uniform-density approximation within cells may be viewed as a lossy compression technique, but we minimize its impact by exactly enforcing Gauss' law. More sophisticated in-cell initialization models (beyond uniform) to improve the fidelity of the spatial distribution can be implemented in the method if so desired.

Since a new set of particles is employed at restart time, their trajectories will be different from the original particles, which will inevitably lead to differences in collective behavior after some time. However, the new set of particles will still represent a valid random ensemble of the reconstructed distribution function, and therefore produce physically correct and relevant dynamics.

\subsection{Performance}

We have performed benchmarks with our implementation (using adaptive component-wise EM) and have concluded that the cost per particle of a single iteration of the the GMM algorithm (without hand-tuned vectorization) is comparable to a single particle push, which makes it competitive with other computations typically performed in PIC codes.  Specifically, we consider a 2D Weibel instability problem on a 16x16 mesh with
two species with 1024 particles (and 64 byte/particle) per species per cell. Fifty time steps
of our implicit code DPIC \cite{chen2015multi}, run serially and
without threading, take 900s, 835 of which are spent on particle pushing.
During that time, $2.2\times10^{9}$ particle pushes are performed,
with about 84 pushes per particle per timestep on average. This gives
$\sim$0.38 $\mu$s/particle-push, or $\sim$32 $\mu$s/particle/$\Delta$t.
The GMM algorithm on the resulting Weibel instability particle set
at $t=50$ (which is fully in the nonlinear stage), using a GMM tolerance
of $10^{-6}$, performs an average of 260 EM iterations per cell taking
49s, resulting in 0.36 $\mu$s/EM-iteration/particle, or 93 $\mu$s/particle
overall. It follows that the unit cost per particle of GMM is about
the same as in PIC (0.36 $\mu$s/EM-iteration/particle vs. 0.38 $\mu$s/particle-push),
and that the cumulative wall-clock time of the GMM reconstruction
is about $3\times$ more expensive than an implicit PIC time step. 

The decompression stage (i.e., reading Gaussian data and sampling particles from them) takes only about $0.3\%$ of the total time, and is therefore negligible.  
Timings have been obtained with one OpenMP thread and one MPI process in a MacBook Pro laptop 2.8 GHz Intel Core i7 with 16 GB 1600 MHz DDR3 memory chips using the gcc 6.5.0 compiler. 

\section{Related work}
Many strategies and techniques have been proposed and studied in the past to mitigate the CR IO bottleneck in large-scale scientific simulations. These include multi-level checkpointing~\cite{moody2010design}, incremental checkpointing~\cite{ferreira2014accelerating}, lazy checkpointing~\cite{tiwari2014lazy}, optimized scheduling~\cite{garg2018shiraz}, checkpointing with data compression~\cite{son2014data,cappello2019use}, etc.
Among these approaches, checkpointing with data compression is the most related to this study. Compression techniques typically fall into two categories: lossless and lossy. Lossless compression techniques are generally inadequate for checkpointing on current/future supercomputers, achieving small (1.1$\times$ to 2$\times$) compression ratios~\cite{son2014data}. Recent advances in lossy compression techniques allow for user-defined error-bounds for the compressed data, which safeguard the compression process. Error-bounded lossy compressors have been explored for CR in various applications (e.g. climate, CFD, chemistry, combustion, etc.) ~\cite{chen2014numarck,sasaki2015exploration,tao2018improving,zhang2019efficient,calhoun2019exploring,reza2019analyzing,triantafyllides2019analyzing,zhang2020bit}, with various levels of success (e.g., compression ratios from a few to several tens or even higher, depending on the error tolerance). For particle-type (N-body) simulations, state-of-art lossy compressors find compression ratios of $\sim2\times$ to $4\times$ with relative error tolerances of $\sim10^{-4}$,  and $4\times$ to $5\times$ with error tolerances of $\sim10^{-2}$~\cite{tao2017depth,cappello2020fulfilling}. However, to the best of our knowledge, no CR capability has been demonstrated for particle simulations. These lossy compressors (e.g.~\cite{lindstrom2014fixed,di2016fast}) employ techniques such as prediction, transforms, quantization, and encoding to directly compress floating-point numbers, which are orthogonal to our approach. Our method, which models the distribution of data and works well for compressing velocity coordinates in particle simulations, can also leverage these other techniques independently (say, to compress the Gaussian data), thereby augmenting its compression potential.
}

\section{Conclusion} 
In summary, we have introduced a high-fidelity, high-compression-ratio technique of
checkpoint/restart for PIC simulations. We employ unsupervised learning of Gaussian mixtures for the particle velocity distribution function, which can significantly reduce CR IO requirements. {\color{black}We have demonstrated compression ratios of about 75, with much larger compression ratios possible for larger number of particles (because the number of Gaussians required to model a given particle distribution does not scale with the number of particles).} Thus, the proposed technique may enable very large-scale particle simulations, which would otherwise be stringently bottlenecked by current and future supercomputer IO systems. 

\pagebreak\bibliographystyle{ieeetr}
\bibliography{GMM}

\end{document}